\documentclass{article}

\usepackage{graphicx}
\usepackage{float}
\usepackage{placeins}
\usepackage[a4paper,margin=1in]{geometry}
\usepackage{amsmath,amssymb}
\usepackage{authblk}
\usepackage[hidelinks]{hyperref}
\usepackage{cite}
\usepackage{enumitem}
\usepackage{caption}
\usepackage{upgreek}
\usepackage{bm}
\DeclareCaptionLabelFormat{figpipe}{\textbf{Fig.~#2\textbar~}}
\title{Primary creep encodes time to failure across laboratory and natural systems}

\author[1*]{Qinghua Lei}
\author[2]{Didier Sornette}

\affil[1]{Department of Earth Sciences, Uppsala University, Uppsala 75236, Sweden}
\affil[2]{Institute of Risk Analysis, Prediction and Management, Academy for Advanced Interdisciplinary Studies, Southern University of Science and Technology, Shenzhen 518055, China}

\date{}

\begin{document}
\maketitle

\begingroup
\renewcommand\thefootnote{}\footnote{
* Correspondence: \texttt{qinghua.lei@geo.uu.se}}
\addtocounter{footnote}{0}
\endgroup

\vspace{-2em}

\begin{abstract}
Geomaterials often exhibit progressive creep characterized by an initial decelerating phase, frequently followed by an extended period of approximately constant deformation rate, and ultimately an accelerating regime leading to catastrophic failure. Despite extensive research, the timing of rupture and its relationship to the different creep phases, particularly in natural systems, remain poorly constrained. Here, we compile creep data from laboratory experiments on rocks, composites, papers, and glasses, together with observations from field systems including landslides, rockfalls, and glaciers. We find that the duration of the early-stage creep, marked by the transition to the minimum (or quasi-stationary) deformation rate, correlates nearly linearly with the time to rupture over five orders of magnitude. This unified scaling highlights that the early-time dynamics reflect the full evolution toward failure, providing a simple and robust framework for forecasting rupture across laboratory and natural systems.
\end{abstract}

\vspace{2em}

\section*{Introduction}

Forecasting the timing of catastrophic failure in geomaterials remains a fundamental challenge in Earth sciences. Laboratory experiments show that materials approaching rupture exhibit characteristic creep behavior, commonly described as three stages \cite{Scholz1968,Brantut2013}: primary creep, with a decelerating strain rate often following Andrade's law \cite{Andrade1910}; secondary creep, with a near-constant strain rate; and tertiary creep, with accelerating deformation leading to failure and often described by time-to-failure power laws \cite{Voight1988,Voight1989,JohansenSornette2000}. However, several experimental and theoretical studies suggest that the secondary stage may not represent a distinct physical regime, but rather a transient crossover between primary and tertiary creeps \cite{Lockner1993,Main2000,AmitranoHelmstetter2006}, which are intrinsically linked \cite{SaichevSornette2005}. Laboratory studies further show that primary creep encodes the time to failure: in heterogeneous materials, the duration of primary creep, corresponding to the time of minimum strain rate, is proportional to the failure time \cite{Nechad2005JMPS,Nechad2005PRL,Koivisto2016}. This proportionality arises because both the relaxation phase and the subsequent acceleration phase are governed by the same underlying dynamics of load redistribution and thermally activated processes in heterogeneous systems \cite{SaichevSornette2005}.

Many gravity-driven mass movements, including landslides, rockfalls, and glacier avalanches, exhibit creep behavior that may culminate in catastrophic failure \cite{Petley2002,Petley2005,Faillettaz2015,Lacroix2020}. However, these signals are often intermittent and influenced by external forcing and internal dynamics \cite{Agliardi2020,Bontemps2020,LeiSornette2025a,LeiSornette2025b,LeiSornette2025c}, obscuring their underlying structure. It therefore remains unclear whether relationships identified in laboratory experiments extend to field-scale phenomena. Here, we examine the link between the duration of primary creep and the time to failure across laboratory and natural systems. We show that they exhibit a near-linear correlation spanning five orders of magnitude in time. This scaling establishes a simple relationship that bridges laboratory measurements and field observations, indicating that primary creep encodes time to failure across scales.

\section*{Results}

We compile creep rupture data from laboratory experiments and field cases reported in the literature, spanning rock, composite, paper, and glass samples as well as landslides, rockfalls, and glaciers. For each case, we analyze the temporal evolution of strain, displacement, or tilt rate. We define a characteristic transition time $t_\mathrm{m}$ as the time at which the deformation rate reaches a minimum. This point marks a change in the evolution of the system, separating an initial decelerating (primary) phase from a subsequent regime of renewed acceleration or quasi-stationary creep \cite{Nechad2005JMPS,Nechad2005PRL,Koivisto2016}. Importantly, the identification of $t_\mathrm{m}$ does not rely on assigning the post-minimum regime to a specific creep phase (e.g., secondary or tertiary), but instead provides a consistent and measurable reference across datasets with varying noise levels and temporal resolutions. The failure time $t_\mathrm{f}$ corresponds to the end of the creep time series, marking system-scale rupture, and is measured relative to the onset of primary creep.

\begin{figure}[htbp]
\centering
\includegraphics[width=\textwidth]{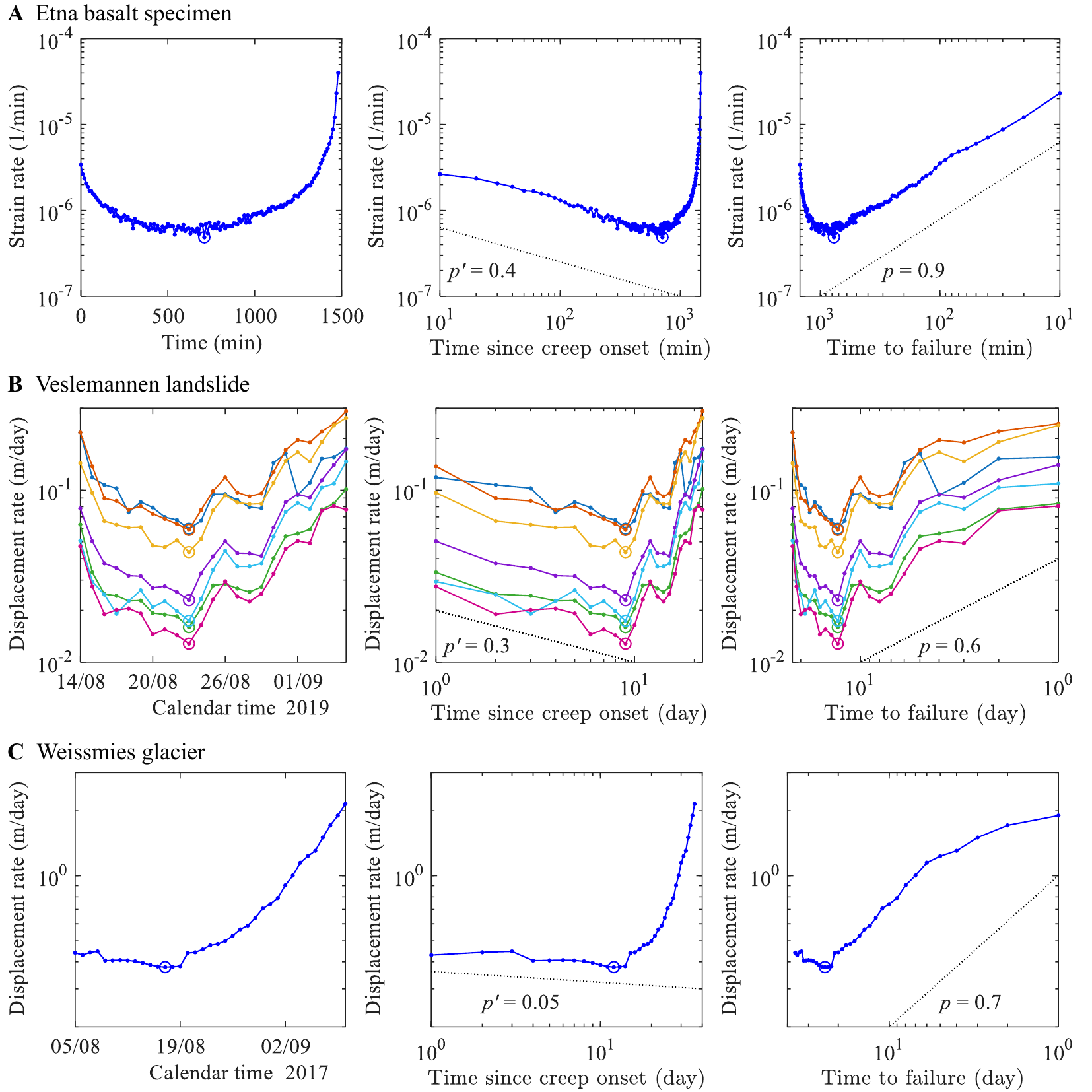}
\caption{Creep evolution in laboratory and natural systems: (A) Etna basalt specimen, (B) Veslemannen landslide, and (C) Weissmies glacier. In each case, creep strain or displacement rate is plotted versus time on a linear scale (left), versus time since the onset of primary creep on a logarithmic scale (middle), and versus time to failure on a logarithmic scale (right). Circles mark the transition time corresponding to the minimum deformation rate. Dashed lines show reference power law trends, with exponents indicated.}
\label{fig:fig1}
\end{figure}

We first show the creep evolution of an Etna basalt specimen \cite{Mansbach2022} subjected to constant load until failure (Fig.~\ref{fig:fig1}A). The raw strain-rate data exhibit high-frequency fluctuations and are therefore smoothed by averaging over successive 10-minute intervals to extract the underlying trend. The creep curve shows an initial decelerating phase, reaching a minimum strain rate at $t_\mathrm{m} = 710$ min, followed by acceleration to failure at $t_\mathrm{f} = 1480$ min (Fig.~\ref{fig:fig1}A, left). When plotted against time since creep onset (Fig.~\ref{fig:fig1}A, middle), the strain rate during primary creep follows an Andrade-type power law decay with exponent $p' \approx 0.4$. When plotted against time to failure (Fig.~\ref{fig:fig1}A, right), the strain rate during tertiary creep exhibits a finite-time singular power law acceleration with exponent $p \approx 0.9$. These results show that both phases follow scaling laws, with the transition time linking the decelerating and accelerating regimes. Similar behavior is observed across a broader set of laboratory creep experiments compiled from published studies and open datasets \cite{Heap2009,Brantut2013,Mansbach2022,Xing2022,Pec2022,BernabePec2022}, spanning basalt, granite, limestone, sandstone, and glass (see Figs.~S1--S4 in Supplemental Material).

The Veslemannen rockslide in western Norway provides a well-documented field case with continuous displacement measurements of slope creep prior to catastrophic failure \cite{Kristensen2021}. Here, we focus on the final creep episode preceding the slope collapse on 5 September 2019 (Fig.~\ref{fig:fig1}B). Displacement rates from seven radar measurement points show consistent temporal evolution, characterized by an initial decelerating phase followed by acceleration to failure (Fig.~\ref{fig:fig1}B, left). The onset of primary creep is defined as the last major velocity peak, occurring on August 14 and associated with a rainfall event on August 13, preceding the final collapse (Fig.~S5). Following this onset time, the displacement rate decreases to a minimum at $t_\mathrm{m} = 9$ days (August 23), after which a persistent acceleration leads to failure at $t_\mathrm{f} = 22$ days (Fig.~\ref{fig:fig1}B, middle). The primary creep follows a power law decay with exponent $p' \approx 0.3$, whereas plotting against time to failure (Fig.~\ref{fig:fig1}B, right) reveals a power law acceleration with exponent $p \approx 0.6$. Despite fluctuations in the time series, the underlying trend remains consistent across all the radar measurement points, with a well-defined transition time linking the decelerating and accelerating regimes. Additional field cases were analyzed, including the Preonzo landslide, the Grabengufer and Gallivaggio rockfalls, and the Weissmies and Planpincieux glaciers \cite{Loew2017,Dematteis2021,Leinauer2023}. Similar creep rupture behavior is observed for these cases, all showing transitions from decelerating to accelerating creep with consistent power law scaling (Fig.~\ref{fig:fig1}C and Fig.~S6).

\begin{figure}[htbp]
\centering
\includegraphics[width=0.9\textwidth]{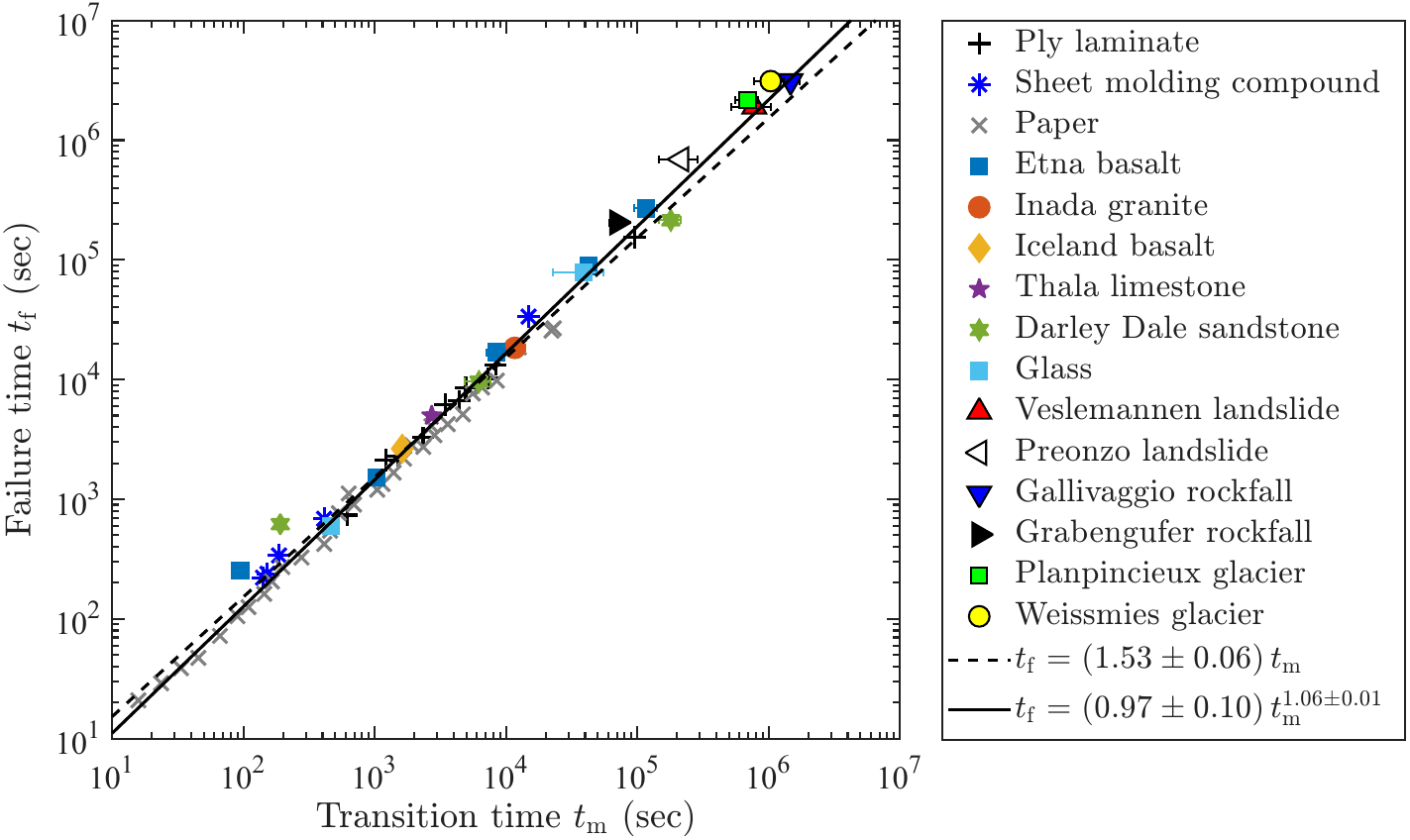}
\caption{Scaling between transition time $t_\mathrm{m}$ and failure time $t_\mathrm{f}$ across laboratory samples (composite, paper, rock, and glass) and natural systems (landslides, rockfalls, and glaciers). The solid line shows a power law fit and the dashed line indicates a linear relation. Error bars represent the uncertainty in transition time, estimated from the local variability of the deformation rate and its temporal gradient at the minimum (see Text~S1 in Supplemental Material).}
\label{fig:fig2}
\end{figure}

We compile the transition time $t_\mathrm{m}$ and failure time $t_\mathrm{f}$ for all laboratory and field cases (Fig.~\ref{fig:fig2}), together with previously published laboratory data on composite materials \cite{Nechad2005JMPS,Nechad2005PRL} and paper sheets \cite{Koivisto2016}. The combined dataset reveals a near-linear correlation between $t_\mathrm{m}$ and $t_\mathrm{f}$ spanning approximately five orders of magnitude in time. A linear fit gives $t_\mathrm{f} = (1.53 \pm 0.06)\, t_\mathrm{m}$, in general consistent with the linear relationships reported for composite specimens \cite{Nechad2005JMPS,Nechad2005PRL} and paper sheets \cite{Koivisto2016}, while extending to rock and glass samples as well as field sites. A power law fit yields $t_\mathrm{f} = (0.97 \pm 0.10)\, t_\mathrm{m}^{1.06 \pm 0.01}$, indicating an exponent close to unity. Despite differences in material properties, loading conditions, environmental forcing, and spatial and temporal scales, the data collapse onto a single trend, demonstrating that the duration of primary creep provides a robust predictor of time to failure across scales.

\FloatBarrier

\section*{Discussion}

The emergence of a single scaling relation across materials, loading conditions, and scales points to a common underlying mechanism controlling the full creep evolution, rather than distinct, unrelated mechanisms for each creep phase. Primary and tertiary creep can be interpreted as successive stages of a single progressive damage process governed by thermal activation and stress redistribution in a heterogeneous medium \cite{Lockner1993,Lockner1998,Main2000,SaichevSornette2005,Nechad2005JMPS}, rather than as independent processes. Primary creep corresponds to a relaxation regime in which deformation preferentially occurs in the weakest regions, progressively exhausting them and producing a power law decay of the deformation rate \cite{WeissAmitrano2023}. As deformation proceeds, thermally activated microcrack growth increases crack density and promotes interactions, leading to damage localization and the onset of tertiary creep that ultimately culminates in catastrophic failure \cite{Lockner1993,Politi2002,AmitranoHelmstetter2006}. The transition between primary and tertiary creep thus marks the progressive localization and collective organization of damage toward macroscopic failure, with early-time relaxation reflecting the distribution of strengths and the efficiency of stress transfer that set the conditions for subsequent damage accumulation and control the time required to reach global instability.

Across the analyzed cases, we systematically observe that exponent $p'$ for primary creep is systematically smaller than exponent $p$ for tertiary creep, revealing an asymmetry between relaxation and acceleration processes. A smaller exponent $p'$ implies a more persistent power law relaxation, meaning that the deformation rate decays more gradually over time than for larger $p'$. This reflects longer effective memory and less efficient stress redistribution in heterogeneous media, where damage remains more spatially distributed. By contrast, a larger value of $p$ corresponds to a more sustained acceleration, with the deformation rate increasing more gradually toward failure, whereas for smaller $p$ the acceleration becomes increasingly concentrated close to rupture. This behavior is consistent with progressive crack interactions and stress redistribution leading to damage localization. Furthermore, both exponents are generally smaller than unity, in contrast to mean-field predictions of values close to one \cite{SaichevSornette2005,Nechad2005JMPS,Nechad2005PRL}. This departure arises from the breakdown of mean-field assumptions, particularly homogeneous stress redistribution, as real systems exhibit spatially variable load transfer, finite-range interactions, and spatially correlated damage evolution.

The observed proportionality between the transition time $t_\mathrm{m}$ and failure time $t_\mathrm{f}$ has important implications for failure prediction. In principle, once a primary creep phase is identified, the time of minimum deformation rate provides an estimate of $t_\mathrm{m}$, from which a conditional forecast of $t_\mathrm{f}$ can be obtained through the near-linear scaling. In natural systems, however, the identification of the onset of primary creep is often ambiguous due to the influence of intermittent external forcing. Strong internal fluctuations or external perturbations can effectively reset the system, generating shocks that define new reference times for subsequent creep evolution (as exemplified by the Veslemannen case; Fig.~S5). Following each such shock, the system undergoes a relaxation phase leading to a rate minimum, from which a tentative prediction can be made. This leads to a prospective forecasting framework in which predictions are continuously updated and occasionally reset, inevitably producing false alarms that are progressively filtered as the system evolves. As failure is approached, the system becomes increasingly governed by internal damage processes, while exogenous perturbations continue to modulate and intermittently reorganize the trajectory toward collapse \cite{LeiSornette2025a}. The nested organization of successive creep episodes further suggests that this evolution reflects not only external triggering but also intrinsic dynamics of damage accumulation, giving rise to an emergent hierarchical temporal structure with scale-invariant characteristics \cite{LeiSornette2025b,LeiSornette2025c}.

\section*{Conclusions}

Our results support the hypothesis that primary and tertiary creep are intrinsically linked stages of a single progressive damage process, and that the duration of primary creep provides a robust predictor of failure time across laboratory and natural systems. The near-linear scaling between transition time and failure time extends laboratory findings to field-scale settings despite differences in material properties, environmental forcing, and spatiotemporal scales. This relationship provides a unifying framework for understanding and forecasting catastrophic failure, bridging controlled experiments and complex natural systems, and offering new opportunities for early warning of geohazards.

\section*{Acknowledgments}

Q.L.\ is grateful for support from the European Research Council (ERC) under the European Union's Horizon Europe programme (ERC Consolidator Grant, grant no.~101232311) for the project ``Unified framework for modelling progressive to catastrophic failure in fractured media (FORECAST)''. Q.L.\ and D.S.\ acknowledge support from Norwegian Water Resources and Energy Directorate for funding the project ``Towards a Next-Generation Landslide Early Warning System''. D.S.\ is grateful for support from the National Natural Science Foundation of China (Grant No.~U2039202, T2350710802) and the Shenzhen Science and Technology Innovation Commission (Grant No.~GJHZ20210705141805017).

\section*{Author Contributions}
Q.L.: Conceptualization, Methodology, Data curation, Formal analysis, Investigation, Visualization, Writing--original draft, Funding acquisition, Project administration. D.S.: Conceptualization, Methodology, Funding acquisition, Writing--review \& editing.

\section*{Preprint Statement}

This manuscript has been submitted to \textit{Geology} and has not yet been peer reviewed by the Geological Society of America.

\vspace{1em}
\noindent\textbf{Supplemental Material.} Text S1 and Figs.~S1--S6 are provided in the supplemental material, available in the TeX source files of the arXiv submission.

\end{document}